\documentstyle[11pt,epsfig]{article}
\def\a{\alpha}

\def\Tr{{\rm Tr}}
\def\dis{\displaystyle}
\def\le{\left(}
\def\ri{\right)}

\def\no{\nonumber}
\def\del{\delta}

\def\G{\Gamma}
\def\rar{\rightarrow}
\def\fra1g2{\frac{1}{g^2}}
\def\dg{{\dagger}}

\def\Kt{\tilde{K}}

\def\e{\epsilon}

\def\f12{\frac{1}{2}}

\def\pd{\partial}
\def\ve{\varepsilon}

\begin{document}
\begin{titlepage}
\flushright{USM-TH-122}\\
\vskip 2cm
\begin{center}
{\Large \bf An approach to solve  Slavnov--Taylor identities \\
\vspace{3mm} in nonsupersymmetric non-Abelian gauge theories} \\
\vskip 1cm  
Igor Kondrashuk, Gorazd Cveti\v{c}, and Iv\'an Schmidt \\
\vskip 5mm  
{\it Departamento de F\'\i sica, Universidad T\'ecnica 
Federico Santa Mar\'\i a, \\
Avenida Espa\~{n}a 1680, Casilla 110-V, Valpara\'\i so, Chile \\
\vskip 5mm 
e-mail:~igor.kondrashuk@fis.utfsm.cl \\
e-mail:~gorazd.cvetic@fis.utfsm.cl \\
e-mail:~ivan.schmidt@fis.utfsm.cl}\\
\end{center}
\vskip 20mm
\begin{abstract}
We present a way to solve Slavnov--Taylor identities in a general 
nonsupersymmetric theory. The solution can be parametrized by 
a limited number of functions of spacetime coordinates, so that 
all the effective fields are dressed by these functions via integral
convolution. The solution restricts the ghost part of the effective action 
and gives predictions for the physical part of the effective action. 
\end{abstract}
\end{titlepage}  

\section{Introduction}

The effective action is an important quantity of the quantum theory. 
Defined as the Legendre transformation of the path integral, it provides 
us with an instrument to find the true vacuum state of the theory under 
consideration 
and to study its behavior taking into account quantum corrections. 
Slavnov--Taylor (ST) identities are also an important tool to prove 
the renormalizability 
of gauge theories in four spacetime dimensions \cite{ST,SF}. They generalize 
Ward--Takahashi identities of quantum electrodynamics to the non-Abelian 
case and can be derived starting from the property of invariance of the 
tree-level action with respect to BRST symmetry \cite{BRST,Becchi}. 
ST identities 
for the effective action  have been derived in Ref. \cite{Lee}.
    
Slavnov--Taylor identities are  equations involving 
variational derivatives of the effective action. The effective action 
contains all the information about quantum behaviour of the theory, 
and in quantum field theory it is the one particle irreducible 
diagram generator. Searching for the solution to Slavnov--Taylor 
identities can be considered as a complementary method to the existing 
nonperturbative methods of quantum field theory such as the Dyson--Swinger 
and Bethe--Salpeter equations.  The solution to Slavnov--Taylor 
identities in the four-dimensional supersymmetric theory has been 
proposed recently \cite{jhep}. In the procedure to derive that solution, 
the no-renormalization theorem for superpotential \cite{GRS,West} 
has been used extensively. In this paper we will suggest that this
point is not crucial and that arguments similar as those 
given before \cite{jhep} 
can be used in the nonsupersymmetric case. In the approach developed 
below there are no restrictions on the number of dimensions and 
renormalizability of the theory. We require only that the theory 
under consideration can be regularized in such a way that the 
Slavnov--Taylor identities are valid and that BRST symmetry is anomaly free,
as in the case, e.g., in QCD.

We argue that the functional structure of the auxiliary ghost-ghost 
$Lc^2 $ correlator in nonsupersymmetric 
gauge theories is fixed by Slavnov--Taylor identities in a unique 
way. In this correlator $L$ is a nonpropagating background field
and it is coupled at the tree level to the 
BRST transformation of the ghost field $c.$ According to our assumption,
the vertex  $Lc^2$ is invariant with respect to ST identities  
and this then gives the following quantum structure for it:  
\begin{eqnarray}
\dis{\int~dx'dxdydz~G_{c}(x'-x)~G^{-1}_{c}(x'-y)
~G^{-1}_{c}(x'-z)\frac{i}{2}f^{bca} L^{a}(x)c^{b}(y)c^{c}(z).}
\label{structure}
\end{eqnarray}
As one can see, the main feature of this result is that 
the effective ghost field $c$ is dressed by the unknown 
function $G^{-1}_{c}(x-y).$ This dressing 
contains all the quantum information about this correlator. 
We can use the structure of this correlator as a starting point to 
find the solution for the total effective action. 
   
The solution to the Slavnov--Taylor identities found in the present 
paper imposes restrictions on the ghost part of the effective 
action. For example, it means that the gluon-ghost-antighost vertex  
can be read off from our  result for the effective action (\ref{G}): 
\begin{eqnarray}
G_m(q,p) = iq_m\frac{\tilde{G}_A(q^2)}{\tilde{G}_A(k^2)
\tilde{G}_c(p^2)},   \label{our}  
\end{eqnarray}
where $\tilde{G}_A$ is the Fourier image of a   
function that dresses gauge field, while $G_m(q,p)$ is the 
gluon-ghost-antighost vertex, $q$ is the  momentum of 
the antighost field $b$ and $p$ is the momentum of the ghost field $c$, 
and $p+k+q=0.$ Another feature of the result obtained here is that 
the physical part of the effective action (\ref{G}) is gauge invariant 
in terms of the effective fields dressed by the dressing functions $G$.
In the result (\ref{G}) for the  effective action information about quantum 
behaviour of the theory is encoded in a {\em finite} number of 
dressing functions and in the running function of the coupling.

The paper is organized in the following way. In Section \ref{s2} 
we review some basic aspects of BRST symmetry and Slavnov--Taylor 
identities for the irreducible vertices. In Section \ref{s3} we show 
how to obtain the functional structure (\ref{structure}) of the 
$Lc^2$ correlator. In Section \ref{s4} we obtain the correlator
linear in another nonpropagating background field $K_m,$ thus 
fixing the terms in the effective action which contain ghost and 
antighost effective fields. In Section \ref{s5} we describe 
higher correlators in $K_m$ and $L.$ In Section \ref{s6} 
we make a conjecture about the form of the physical (pure gluonic) 
part of the effective action and then in Section \ref{s7} 
we consider renormalization of it to remove infinities.
A brief summary is given at the end.  The questions of consistency of 
this effective action within perturbative QCD are investigated in 
a second paper \cite{second}. For simplicity, in the present 
paper we focus on  pure gauge theories in four spacetime dimensions 
with $SU(N)$ gauge group. No matter field is included in the 
consideration, although their addition does not change our 
results.

\section{Preliminaries}  \label{s2}

We consider the traditional Yang--Mills Lagrangian of the pure  
gauge theory        
\begin{eqnarray}
 \dis{S = -\int dx~\frac{1}{2g^2}~\Tr\left[F_{mn}(x)F_{mn}(x)\right]}
\label{theory}
\end{eqnarray}

The gauge field is in the adjoint representation of the gauge group. 
A nonlinear local (gauge) transformation of the gauge fields exists 
which keeps 
theory (\ref{theory}) invariant. This symmetry must be fixed, Faddeev--Popov
ghost fields \cite{FP} must be introduced and finally the BRST symmetry 
can be established  for the theory that in addition to the classical 
action (\ref{theory}) contains a Faddeev--Popov ghost action and the 
gauge-fixing term.     

To be specific, we choose the Lorentz gauge fixing condition 
\begin{eqnarray} 
\pd_m A_m(x) = f(x).  \label{GF}
\end{eqnarray} 
Here $f$ is an arbitrary function in the adjoint representation of the 
gauge group 
that is independent on the gauge field. The normalization of the gauge 
group generators is 
\begin{eqnarray*}
\Tr~ \le T^aT^b \ri = \f12\delta^{ab}, ~~~  \le T^a \ri^\dg = T^a, ~~~
\left[T^b,T^c\right] = if^{bca}T^a,
\end{eqnarray*}  
and we use notation $X = X^aT^a$ for all the fields in the adjoint 
representation of the gauge group, like gauge fields themselves, ghost fields, 
and their respective sources.

The conventional averaging procedure with respect to $f$ is applied 
to the path integral with the weight
\begin{eqnarray*}
e^{~\dis{-i~\int~d~x~~\Tr~\frac{1}{\a}~f^2(x)}}
\end{eqnarray*}   
and as the result we obtain the path integral 
\begin{eqnarray}
& \dis{Z[J,\eta,\rho,K,L] = \int~
  dA~dc~db~\exp i} \Bigg\{  \dis{S[A,b,c]}
  \Biggr.  \no \\
& \Biggl. + \dis{2~\Tr\le\int~dx~J_m(x)A_m(x) + i\int~dx~\eta(x) c(x)
  + i\int~dx~\rho(x) b(x)\ri} \Biggr. \label{pathRMnon}\\
& \Biggl. + \dis{2~\Tr\le i\int~dx~K_m(x)\nabla_m~c(x) + \int~dx~L(x)c^2(x)\ri }
  \Bigg\} . \no
\end{eqnarray}
in which 
\begin{eqnarray}
& S[A,b,c] = \dis{\int~d~x~\left[-\frac{1}{2g^2}~\Tr\left[F_{mn}(x)F_{mn}(x)\right] 
    - ~\Tr\le\frac{1}{\a}\left[\pd_m A_m(x)\right]^2 \ri  \right.} \no\\ 
& - \dis{\left.    2~\Tr~\le~i~b(x)\pd_m~\nabla_m~c(x)\ri \right]}.  \label{action}
\end{eqnarray} 
Here the ghost field $c$ and the antighost field $b$ are Hermitian, 
$b^\dg = b$, $c^\dg = c$ in the adjoint representation of the gauge 
group. They possess Fermi statistics. 

The infinitesimal transformation of the gauge field $A_m$ is defined
by the fact that it is a gauge connection,
\begin{eqnarray*}
A_m \rar  A_m - \nabla_m \lambda,    
\end{eqnarray*} 
where $\lambda(x)$ is an infinitesimal parameter of the gauge 
transformation. This transformation comes from the transformation of 
covariant derivatives, 
\begin{eqnarray*}
\nabla_m \rar e^{i\lambda}~\nabla_m~e^{-i\lambda}, 
~~~ \nabla_m = \pd_m +i A_m, ~~~\phi \rar  e^{i\lambda}~\phi,
\end{eqnarray*}  
where $\phi$ is some representation of the gauge group.
To obtain the BRST symmetry we have to substitute $i~c(x)~\ve$ instead 
of $\lambda.$ Here $\ve$ is Hermitian Grassmannian parameter, 
$\ve^\dg = \ve,$ $\ve^2 =0.$ Thus, the BRST transformation of 
the gauge field is 
\begin{eqnarray}
A_m \rar A_m - i~\nabla_m c~\ve.  \label{BRST1}
\end{eqnarray} 
In order to obtain the BRST transformation of the ghost field $c$ 
we have to consider two subsequent BRST transformations
\begin{eqnarray}
\nabla_m \rar \dis{e^{-c\kappa}e^{-c\ve}~\nabla_m~e^{c\ve}e^{c\kappa} = 
e^{-c\ve-c\kappa-(c\ve)(c\kappa)}~\nabla_m~
e^{c\ve+c\kappa+(c\ve)(c\kappa)}},  \label{BRSTtrick} 
\end{eqnarray}  
where $\kappa$ is a Grassmannian parameter too, $\kappa^2=0.$ This transformation 
again is equivalent to an infinitesimal transformation of the gauge field 
in covariant derivatives,
\begin{eqnarray*}
A_m \rar  A_m -  i\nabla_m ~\left[c\ve+c\kappa +(c\ve)(c\kappa)\right].
\end{eqnarray*} 
It means that we can consider the inner BRST transformation (with $\ve$) as 
the substitution (\ref{BRST1}) in the outer BRST transformation (with $\kappa$). 
The second term after the covariant derivative is a transformation 
of $A_m$ under the outer BRST transformation while the third term after the 
covariant derivative is the transformation of $i\nabla_m ~c\kappa$ 
and can it be cancelled by the transformation of the second term $c\kappa$  
\begin{eqnarray}
c \rar   c + c^2\ve.   \label{BRST2}
\end{eqnarray}    
Thus, the transformations (\ref{BRST1}) and (\ref{BRST2}) together leave the 
covariant derivative of the ghost field unchanged. Such a symmetry is very 
general and always exists if the gauge fixing procedure has been performed
in the path integral for any theory with nonlinear local symmetry. The 
noninvariance of the gauge fixing term is cancelled by the corresponding 
transformation of the antighost field $b.$

To collect all things together, the action (\ref{action}) is invariant with 
respect to the BRST symmetry transformation with Grassmannian parameter 
$\ve,$
\begin{eqnarray}  
& A_m \rar A_m - i~\nabla_m c~\ve, \no\\
& c \rar  c + c^2\ve, \label{BRSTall}\\
& \dis{b \rar  b - \frac{1}{\a}~\pd_m~A_m}\ve. \no
\end{eqnarray}

The external sources $K$ and $L$ of the BRST transformations of the 
fields are BRST invariant by definition, so the last two lines in the
Eq. (\ref{pathRMnon}) are BRST invariant with respect to the transformations
(\ref{BRSTall}).

The effective action $\G$ is related to $W = i~ln~Z$ by the Legendre
transformation\footnote{We have traditionally  used in this paper
the same notation for variable of the effective action and for 
variable of integration in the path integral coupled to the corresponding 
source \cite{SF}.} 
 \begin{eqnarray}
& \dis{A_m \equiv - \frac{\del W}{\del J_m}, 
      ~~ ic  \equiv - \frac{\del W}{\del \eta}, 
      ~~ ib \equiv - \frac{\del W}{\del \rho}}  \label{defphi} \\
& \dis{\G = - W - 2~\Tr\le \int~d~x~J_m(x)~A_m(x) + \int~d~x~i\eta(x)~c(x)
                + \int~d~x~i\rho(x)~b(x) \ri } \no\\
&  \equiv \dis{- W - 2~\Tr\le X\varphi\ri,} \label{Legendre} \\
&  \dis{\le X\varphi\ri \equiv i^{G(k)}\le X^{k}\varphi^{k}\ri,}  \no\\
& \dis{X \equiv \le J_m,\eta,\rho\ri,  ~~~   \varphi \equiv
  \le A_m,c,b\ri}. \no
\end{eqnarray}
where $G(k) = 0$ if $\varphi^{k}$ is Bose field and $G(k) = 1$
if $\varphi^{k}$ is Fermi field. We use throughout the paper 
notation  
\begin{eqnarray*}
\frac{\del}{\del X} = T^a\frac{\del}{\del X^a}
\end{eqnarray*}
for any field $X$ in the adjoint representation of the gauge group.
Iteratively all equations (\ref{defphi})
can be reversed,
\begin{eqnarray*}
X = X[\varphi,K_m,L]
\end{eqnarray*}
and the effective action is defined in terms of new variables,
$\G = \G[\varphi,K_m,L].$
Hence, the following equalities take place
\begin{eqnarray}
 \dis{\frac{\del \G}{\del A_m} = - J_m, ~~~ 
  \frac{\del \G}{\del K_m} = - \frac{\del W}{\del K_m},~~~
 \frac{\del \G}{\del c} = i\eta, ~~~ 
\frac{\del \G}{\del b} = i\rho, ~~~
\frac{\del \G}{\del L} = - \frac{\del W}{\del L}}. \label{GW}
\end{eqnarray}

If the change of fields (\ref{BRSTall}) in the path integral (\ref{pathRMnon}) is made
one obtains the Slavnov--Taylor identity as the result of invariance of the
integral (\ref{pathRMnon}) under a change of variables,
\begin{eqnarray}
& \Tr\left[\dis{\int dx J_m(x)\frac{\del}{\del K_m(x)} 
  - \int dx i\eta(x)\le\frac{\del}{\del L(x)}\ri} \right. \no\\
&  + \left.\dis{ \int dx i\rho(x)\le\frac{1}{\a}\pd_m\frac{\del}
{\del J_m(x)}\ri}\right]W = 0,
\label{STprom}
\end{eqnarray}
or, taking into account the relations (\ref{GW}), we have \cite{SF}
\begin{eqnarray}
& \Tr\left[\dis{\int~d~x~\frac{\del \G}{\del A_m(x)}\frac{\del \G}{\del K_m(x)}
  + \int~d~x~\frac{\del \G}{\del c(x)}\frac{\del \G}{\del L(x)}} \right. \no\\
&  \left. - \dis{\int~d~x~\frac{\del \G}{\del b(x)}
   \le\frac{1}{\a}~\pd_m~A_m(x)\ri} \right] = 0. \label{STrMnon} 
\end{eqnarray}

The problem is to find the most general functional $\G$ of the variables
$\varphi,K_m,L$ that satisfies the ST identity (\ref{STrMnon}). 
Before doing it, we need in addition 
to ST identities also the ghost equation
that can be derived by shifting the antighost field $b$ by an arbitrary 
field $\ve(x)$ in the path integral (\ref{pathRMnon}). The consequence of 
invariance of the path integral with respect to such a change of 
variable is (in terms of the variables (\ref{defphi})) \cite{SF}
\begin{eqnarray}
\frac{\del \G}{\del b(x)} + \pd_m~\frac{\del \G}{\del K_m(x)} = 0. \label{ghost}
\end{eqnarray}
The ghost equation (\ref{ghost}) restricts 
the dependence of $\G$ on the antighost field $b$ and on the external 
source $K_m$ to an arbitrary dependence on their combination 
\begin{eqnarray}
\pd_m~b(x) + K_m(x). \label{comb}
\end{eqnarray}
This equation together with the third term in the ST identities 
(\ref{STrMnon}) is responsible for the absence of quantum corrections to the 
gauge fixing term. Stated otherwise, when expressing 
$\del \G/\del b(x)$ in the third term in the ST identity
(\ref{STrMnon}) as $- \pd_m ( \del \G/\del K_m(x))$
by Eq.~(\ref{ghost}), the sum of the first and the third term
in (\ref{STrMnon}) can be rewritten as
\begin{eqnarray*}
{\rm  Tr} \int dx  
\dis{\frac{\del  \G' }{\del A_m(x)}
\frac{\del  \G' }{\del K_m(x)}},
\end{eqnarray*}
where $\G' \equiv \Gamma - S^{\rm (g.f.)}$, and
$\dis{S^{\rm (g.f.)} = - (1/\a) {\rm Tr} \int dx  [ \pd_m A_m(x)]^2}$
is the gauge--fixing part of the classical action (\ref{action}).
In fact, all the other terms in the ST identity (\ref{STrMnon})
can be rewritten with $ \G'$ instead of $\G$,
yielding
\begin{eqnarray}
& \Tr\left[\dis{\int~d~x~\frac{\del \G'}{\del A_m(x)}\frac{\del \G'}{\del K_m(x)}
  + \int~d~x~\frac{\del \G'}{\del c(x)}\frac{\del \G'}{\del L(x)}} \right] = 0.
\label{STred} 
\end{eqnarray}
This shows explicitly that the gauge--fixing part of $\Gamma$
remains unaffected by quantum corrections
($ \G =  \G' + \G^{\rm (g.f.)}$; $\G^{\rm (g.f.)} = S^{\rm (g.f.)}$).

\section{Functional structure of $Lcc$ vertex} \label{s3}

One can consider the part of the effective action that depends only 
on the fields $L$ and $c.$ We write generally   
\begin{eqnarray}
& \dis{\G|_{L,c} = \int~dx_1~dy_1~dy_2~\G|^{(a_1;b_1,b_2)}_{L,c}(x_1;y_1,y_2)
   L^{a_1}(x_1)c^{b_1}(y_1)c^{b_2}(y_2)} + \dots \no\\ 
& \dots + \dis{\int~dx_1\dots~dx_n~dy_1\dots~dy_{2n}~
 \G|^{(a_1,\dots,a_n;b_1,\dots,b_{2n})}_{L,c}(x_1,
   \dots,x_n;y_1,\dots,y_{2n})
 ~L^{a_1}(x_1)} \times\no\\ 
& \dis{\times \dots L^{a_n}(x_n)c^{b_1}(y_1)\dots c^{b_{2n}}(y_{2n})} + 
  \dots  \label{dec}
\end{eqnarray}

We assume that the first term is invariant with respect to 
the second operator in the identities (\ref{STrMnon})
which is 
\begin{eqnarray}
& \dis{\Tr~\int~d~x~\frac{\del \G}{\del c(x)}
     \frac{\del \G}{\del L(x)} = 0}. \label{STrMnonsmall}
\end{eqnarray}
This assumption is based on the following. In  perturbation theory 
the first term of (\ref{dec}) can be understood as the classical term plus 
a quantum correction to the vertex  $Lcc$ (nothing forbids us to consider 
the auxiliary field $L$ as a non-propagating background field). The operator 
(\ref{STrMnonsmall}) can be considered as an infinitesimal substitution 
in the effective action 
\begin{eqnarray}
c(x) \rar c(x) + \frac{\del \G}{\del L(x)}.  \label{aux4}
\end{eqnarray}
In other words, one can consider the result of such a substitution 
as the difference  
\begin{eqnarray*}
\G\left[ L,c(x)+\frac{\del\G}{\del L(x)}\right] -
\G\left[ L,c(x)\right], 
\end{eqnarray*}
to linear order in $\dis{\frac{\del\G}{\del L(x)}}.$ As one can see, 
the application of the substitution (\ref{aux4}) to the vertex $Lcc$ 
of the effective action $\G$ gives a variation of order $Lccc$. 
Another contribution of the same order $Lccc$ comes into the variation 
from the monomial  $LccA$ of the effective action $\G$ due to the 
first term in the ST identity (\ref{STrMnon}). Indeed, one can consider     
the first term in  (\ref{STrMnon}) as the substitution
\begin{eqnarray*}
A_m(x) \rar A_m(x) 
+ \frac{\del \G}{\del K_m(x)}, 
\end{eqnarray*}
or, in other words, such a substitution can be considered as the difference  
\begin{eqnarray*}
\G\left[ K_m, A_m(x)+\frac{\del\G}{\del K_m(x)}\right] 
- \G\left[ K_m, A_m(x)\right] 
\end{eqnarray*}
to linear order in $\dis{\frac{\del\G}{\del K_m(x)}}.$ Application 
of such a substitution to the monomial  $LccA$ of the effective action $\G$
gives a contribution of order $Lccc$ in effective fields and this 
contribution comes from full ghost propagator of order in fields $K_m~\pd_m~c,$
\begin{eqnarray*}
LccA \rar Lcc\frac{\del \G}{\del K_m} \sim  
Lcc \frac{\del \le K_m~\pd_m c\ri }{\del K_m } \sim Lccc.
\end{eqnarray*}
Thus, there are only these two possible contributions in variation $Lccc.$ 
Schematically, total $Lccc$ variation can be presented as 
\begin{eqnarray}
\left<Lcc\right> \times \left<Lcc\right> + \left<LccA\right> 
\times \left<K_m\pd_mc\right> = 0  \label{rough}
\end{eqnarray}
where brackets mean v.e.v.s of the vertices. This is a schematic form of the 
ST identity relating the $Lcc$ and $LccA$ field monomials. The precise form 
of this relation can be obtain by differentiating the identity (\ref{STrMnon}) 
with respect to $L$ and three times with respect to $c$ and then by 
setting all the variables of the effective action to zero. The brackets in   
(\ref{rough}) mean that we have taken functional derivatives with respect to 
fields in the corresponding brackets and then have put all the 
effective fields to zero.  
Of course, this sum (\ref{rough}) should be zero since on the r.h.s. of the ST 
identity (\ref{STrMnon}) we have zero. 
One can consider the identity (\ref{rough}) order by order in $g^2$. At tree 
level, the second contribution is absent since  the 
$LccA$ term is absent in the 
classical action. For the first one we obtain the Jacobi identity. At one loop 
level, we have one loop $Lcc$ times tree level $Lcc$ plus one loop $LccA$ 
times tree level $K\pd c.$ However, one loop $LccA$ is superficially 
convergent and does not depend on the normalization point $\mu.$ In 
the asymptotic region one loop  $Lcc$ depends on first degree of 
$\ln (p^2/\mu^2)$ 
where we have taken the symmetric point in momentum space, that is all 
the external momentums of the vertex $Lcc$ are $\sim p^2.$ This means 
that the first degree of $\ln (p^2/\mu^2)$ in one loop $Lcc$ is 
invariant with respect to the operator (\ref{STrMnonsmall}). 
In other words, the dependence 
on  $\ln (p^2/\mu^2)$ is cancelled within the first term of 
the identity (\ref{rough}). We can consider two loop approximation
for the identity (\ref{rough}) in the same manner. Indeed, at  
two loop level of the identity (\ref{rough}) one has two loop   
$Lcc$ times tree level $Lcc$ plus one loop $Lcc$ times one loop $Lcc$
plus two loop $LccA$ times tree level $K\pd c$ plus one loop 
$LccA$ times one loop $K\pd c$ and all this should be zero. However, 
one can see that the second degree of  $\ln (p^2/\mu^2)$ is 
determined again by only the first term in the schematic 
identity (\ref{rough}) 
since two loop  $LccA$ does not have superficial divergences and is 
divergent only in subgraphs. Thus, the second degree of $\ln (p^2/\mu^2)$
is also determined by the invariance with respect to the first 
term in the identity (\ref{rough}). We can go further in this 
logical chain and we will always conclude that the highest degree 
of $\ln (p^2/\mu^2)$ in $Lcc$ is invariant itself with respect 
to ST identity. This is the main source of the intuitive motivation  
to consider the $Lcc$ correlator separately from the other field monomial 
$LccA.$   
 
In such a case, it will be shown below that the only solution for 
this $Lcc$ term of the effective action is  
\begin{eqnarray}
\int dxdx_1dy_1dy_2~G_{c}(x-x_1)~G^{-1}_{c}(x-y_1)
     ~G^{-1}_{c}(x-y_2)~2\Tr\le L(x_1)c(y_1)c(y_2)\ri. 
\label{firstLc^2}
\end{eqnarray}

To prove (\ref{firstLc^2}), we consider the proper correlator 
\begin{eqnarray}
\dis{\G = \int~dx~dy~dz~\G (x,y,z)T^{abc}L^{a}(x)c^{b}(y)c^{c}(z).} 
\label{A1}
\end{eqnarray}
As we have  already noted, in perturbation theory 
it can be understood as a correction to the vertex  $Lc^2$ and we consider  
the auxiliary field $L$ as a non-propagating background field. 
$T^{abc}$ is some group structure. The equation (\ref{A1}) is just a general 
parametrization of the proper correlator $Lc^2$  and nothing more.  
Equation (\ref{A1}) says that 
$\G^{(a;b,c)}(x,y,z) = \G (x,y,z)T^{abc},$
where $T^{abc}$ is a 3-tensor in the adjoint representation of the 
gauge group. This reflects the fact that the global symmetry 
of the gauge group must be conserved in the effective action. 
With respect to that symmetry the auxiliary fields $K^a$ and $L^a$ are 
vectors in the adjoint representation of the gauge group. Also, 
\begin{eqnarray}
\dis{\G (x,y,z)T^{abc} = - \G (x,z,y)T^{acb}.} 
\label{A100}
\end{eqnarray}
This is a direct consequence of the Grassmannian nature of the ghost fields.
It follows from the parametrization (\ref{A1}). Further,
from Eq.  (\ref{A1}) follows
\begin{eqnarray*}
\frac{\del \G}{\del L^a(x)} = \int~dy~dz~\G (x,y,z)~T^{abc}c^b(y)c^c(z). 
\end{eqnarray*}
By substituting this expression in the Slavnov--Taylor identity (\ref{STrMnon})
we have 
\begin{eqnarray*}
& \dis{\int~dx~\frac{\del \G}{\del c^a(x)} \frac{\del \G}{\del L^a(x)} 
 = \int~dx~dy'~dz'~\G (y',x,z')T^{dab}L^d(y')
\frac{\del \G}{\del L^a(x)}c^b(z')}\\
& \dis{ - \int~dx~dy'~dz'~\G (y',z',x)T^{dba}L^d(y')c^b(z')
\frac{\del \G}{\del L^a(x)}}  \\ 
& = \dis{\int~dx~dy~dz~dy'~dz'~\G (y',x,z')T^{dab}L^d(y')\G (x,y,z)T^{amn}
 c^m(y)c^n(z)c^b(z')} \\ 
& - \dis{\int~dx~dy~dz~dy'~dz'~\G (y',z',x)T^{dba}L^d(y')c^b(z')\G (x,y,z)T^{amn}c^m(y)c^n(z)} \\
& = \dis{\int~dx~dy~dz~dy'~dz'~\G (y',x,z')\G (x,y,z)T^{dab}T^{amn}L^d(y')
 c^m(y)c^n(z)c^b(z')} \\ 
& - \dis{\int~dx~dy~dz~dy'~dz'~\G (y',y,x)\G (x,z,z')T^{dma}T^{anb}L^d(y')
 c^m(y)c^n(z)c^b(z')} \\ 
& = \dis{\int~dx~dy~dz~dy'~dz'~\left[\G (y',x,z')\G (x,y,z)T^{dab}T^{amn} 
\right.} \\
& - \dis{\left .\G (y',y,x)\G (x,z,z')T^{dma}T^{anb}\right] 
L^d(y')c^m(y)c^n(z)c^b(z')} = 0.
\end{eqnarray*}
Taking into account (\ref{A100}) the last two lines can be re-written as 
\begin{eqnarray*}
& \dis{\int~dx~dy~dz~dy'~dz'~\left[\G (y',x,z')\G (x,y,z)T^{dab}T^{amn} 
\right.} \\
& - \dis{\left .\G (y',y,x)\G (x,z,z')T^{dma}T^{anb}\right] 
L^d(y')c^m(y)c^n(z)c^b(z')} \\
& = \dis{\int~dx~dy~dz~dy'~dz'~\left[\G (y',x,z')\G (x,y,z)T^{dab}T^{amn} 
\right.} \\
& - \dis{\left .\G (y',x,y)\G (x,z',z)T^{dam}T^{abn}\right] 
L^d(y')c^m(y)c^n(z)c^b(z')} \\
& = \dis{2\int~dx~dy~dz~dy'~dz'~\G (y',x,z')\G (x,y,z)T^{dab}T^{amn}
    L^d(y')c^m(y)c^n(z)c^b(z')} = 0. 
\end{eqnarray*}
Now one can make total symmetrisation with respect to pairs $(m,y)$, 
$(n,z)$, and $(b,z').$ It results in 
\begin{eqnarray*}
& \dis{\int~dx~dy~dz~dy'~dz'~\left[\G (y',x,z')\G (x,y,z)T^{dab}T^{amn} 
  + \G (y',x,y)\G (x,z,z')T^{dam}T^{anb} \right. } \no\\
& \dis{+ \left. \G (y',x,z)\G (x,z',y)T^{dan}T^{abm} \right]
L^d(y')c^m(y)c^n(z)c^b(z') = 0.} 
\end{eqnarray*}
Thus, one comes to the equation
\begin{eqnarray}
& \dis{\int~dx~\G (y',x,z')\G (x,y,z)T^{dab}T^{amn} +\int~dx~ \G (y',x,y)
\G (x,z,z')T^{dam}T^{anb}  } \no\\
& + \dis{ \int~dx~\G (y',x,z)\G (x,z',y)T^{dan}T^{abm}  = 0.} 
\label{identitynew}
\end{eqnarray}
As one can see, at tree level $T^{dab} \sim f^{abd}$ and   
\begin{eqnarray}
\G_{tree} (x,y,z) = \int~dx'\del(x'-x)\del(x'-y)\del(x'-z) 
\label{tree}
\end{eqnarray}
and, hence, the identity (\ref{identitynew}) is Jacobi identity.
We consider in this paper gauge theories with $SU(N)$ gauge group and 
we have noted this in Introduction. The structure constants $f^{abc}$ 
are completely antisymmetric in such a case. With help of identities 
\begin{eqnarray*}
f^{ABC}f^{CDE}f^{EBF} = -\frac{1}{2}~N~f^{ABF}  
\end{eqnarray*}
which are consequences of Jacobi identity, one can reduce the group 
structure of one loop diagram $Lcc$ to $f^{ABC}$ and that is true for 
all loops. Thus, it is natural to assume that
$T^{abc} \sim f^{bca}$ and the identity (\ref{identitynew})
is 
\begin{eqnarray*}
& \dis{\int~dx~\G (y',x,z')\G (x,y,z)f^{abd}f^{mna} 
  + \int~dx~\G (y',x,y)\G (x,z,z')f^{amd}f^{nba}  } \\
& \dis{+ \int~dx~\G (y',x,z)\G (x,z',y)f^{and}f^{bma}  = 0.} 
\end{eqnarray*}
Because of Jacobi identity only two group structures are 
independent here:
\begin{eqnarray*}
& \dis{\left[\int~dx~\G (y',x,z')\G (x,y,z) - 
\int~dx~\G (y',x,y)\G (x,z,z')\right]
 f^{abd}f^{mna}}  \\
& + \dis{\left[ \int~dx~\G (y',x,z)\G (x,z',y) -
\int~dx~ \G (y',x,y)\G (x,z,z')\right] f^{and}f^{bma}  = 0.} 
\end{eqnarray*}
Since these two group structures are independent, we come to the equations
\begin{eqnarray}
& \dis{\int~dx~\G (y',x,z')\G (x,y,z) = \int~dx~\G (y',y,x)\G (x,z,z')} \no\\
& = \dis{\int~dx~ \G (y',x,z)\G (x,z',y).}  \label{equalities}
\end{eqnarray}
We can solve at the beginning the first one:
\begin{eqnarray}
\dis{\int~dx~\G (y',x,z')\G (x,y,z) = \int~dx~\G (y',y,x)\G (x,z,z')}
\label{cond1} 
\end{eqnarray}
and then to check that the second equality is also satisfied. In writing 
this equation we have used the symmetry properties (\ref{A100}). 
We introduce Fourier transformations\footnote{We do not write factors 
$2\pi$ in these Fourier transformations since at the end of the 
calculations we will go back to 
coordinate space in which all the factors $2\pi$ will disappear.}  
\begin{eqnarray*}
& \dis{\G (x,y,z) = \int~dp_1~dq_1~dk_1\del(p_1+q_1+k_1)\tilde{\G}(p_1,q_1,k_1)
\exp(ip_1x +iq_1y + ik_1z)} \\
& \dis{\G (y',x,z') = \int~dp_2~dq_2~dk_2\del(p_2+q_2+k_2)
\tilde{\G}(p_2,q_2,k_2)
\exp(ip_2y' +iq_2x + ik_2z')} \\
& \dis{\G (y',y,x) = \int~dp_3~dq_3~dk_3\del(p_3+q_3+k_3)
\tilde{\G}(p_3,q_3,k_3)
\exp(ip_3y' +iq_3y + ik_3x)} \\
& \dis{\G (x,z,z') = \int~dp_4~dq_4~dk_4\del(p_4+q_4+k_4)
\tilde{\G}(p_4,q_4,k_4)\exp(ip_4x +iq_4z + ik_4z')} 
\end{eqnarray*}
The condition (\ref{cond1}) in momentum space is 
\begin{eqnarray*}
& \dis{\int~dxdp_1dq_1dk_1dp_2dq_2dk_2~\del(p_1+q_1+k_1)\del(p_2+q_2+k_2)
\tilde{\G}(p_1,q_1,k_1)\times} \\
& \dis{\times\tilde{\G}(p_2,q_2,k_2)\exp(ip_1x +iq_1y + ik_1z+ ip_2y' +iq_2x 
+ ik_2z')} \\
& = \dis{\int~dxdp_3dq_3dk_3dp_4dq_4dk_4~\del(p_3+q_3+k_3)\del(p_4+q_4+k_4)
\tilde{\G}(p_3,q_3,k_3)\times} \\
& \dis{\times\tilde{\G}(p_4,q_4,k_4)\exp(ip_3y' +iq_3y + ik_3x+ ip_4x +iq_4z 
+ ik_4z').}
\end{eqnarray*}
It can be transformed to 
\begin{eqnarray*}
& \dis{\int~dp_1dq_1dk_1dp_2dk_2~\del(p_1+q_1+k_1)\del(p_2-p_1+k_2)
\tilde{\G}(p_1,q_1,k_1)\times} \\
& \dis{\times\tilde{\G}(p_2,-p_1,k_2)\exp(iq_1y + ik_1z+ ip_2y' + ik_2z')} \\
& = \dis{\int~dp_3dq_3dk_3dq_4dk_4~\del(p_3+q_3+k_3)\del(-k_3+q_4+k_4)
\tilde{\G}(p_3,q_3,k_3)\times} \\
& \dis{\times\tilde{\G}(-k_3,q_4,k_4)\exp(ip_3y' +iq_3y +iq_4z + ik_4z'),}
\end{eqnarray*}
and then by momentum redefinitions in the second integral one obtains  
\begin{eqnarray*}
& \dis{\int~dp_1dq_1dk_1dp_2dk_2~\del(p_1+q_1+k_1)\del(p_2-p_1+k_2)
\tilde{\G}(p_1,q_1,k_1)\times} \\
& \dis{\times\tilde{\G}(p_2,-p_1,k_2)\exp(iq_1y + ik_1z+ ip_2y' + ik_2z')} \\
& = \dis{\int~dp_2dq_1dk_3dk_1dk_2~\del(p_2+q_1+k_3)\del(-k_3+k_1+k_2)
\tilde{\G}(p_2,q_1,k_3)\times} \\
& \dis{\times\tilde{\G}(-k_3,k_1,k_2)\exp(iq_1y + ik_1z+ ip_2y' + ik_2z').}
\end{eqnarray*}
By removing one of delta functions in each part one obtains 
\begin{eqnarray*}
& \dis{\int~dq_1dk_1dp_2dk_2~\del(p_2+k_2+q_1+k_1)
\tilde{\G}(p_2+k_2,q_1,k_1)\tilde{\G}(p_2,-p_2-k_2,k_2)\times} \\
& \dis{\times\exp(iq_1y + ik_1z+ ip_2y' + ik_2z')} \\
& = \dis{\int~dp_2dq_1dk_1dk_2~\del(p_2+q_1+k_1+k_2)
\tilde{\G}(p_2,q_1,k_1+k_2)\tilde{\G}(-k_1-k_2,k_1,k_2)\times} \\
& \dis{\times\exp(iq_1y + ik_1z+ ip_2y' + ik_2z').}
\end{eqnarray*}
By making the last simplification one obtains 
\begin{eqnarray*}
& \dis{\int~dk_1dp_2dk_2~
\tilde{\G}(p_2+k_2,-p_2-k_2-k_1,k_1)\tilde{\G}(p_2,-p_2-k_2,k_2)\times} \\
& \dis{\times\exp(i(-p_2-k_2-k_1)y + ik_1z+ ip_2y' + ik_2z')} \\
& = \dis{\int~dp_2dk_1dk_2~
\tilde{\G}(p_2,-p_2-k_2-k_1,k_1+k_2)\tilde{\G}(-k_1-k_2,k_1,k_2)\times} \\
& \dis{\times\exp(i(-p_2-k_2-k_1)y + ik_1z+ ip_2y' + ik_2z').}
\end{eqnarray*}
Thus, finally the condition (\ref{cond1}) takes the form 
\begin{eqnarray}
& \dis{\tilde{\G}(p_2+k_2,-p_2-k_2-k_1,k_1)
\tilde{\G}(p_2,-p_2-k_2,k_2)} \label{basic}\\
& \dis{=\tilde{\G}(p_2,-p_2-k_2-k_1,k_1+k_2)
\tilde{\G}(-k_1-k_2,k_1,k_2)}. \no
\end{eqnarray}
This is an equation for a function of three variables, which will be  solved 
below. First we show   that there is simple  
ansatz which satisfies  Eq. (\ref{basic}). Indeed, by choosing  
ansatz 
\begin{eqnarray}
\dis{\tilde{\G}(p,q,k) = \frac{\tilde{G}(q^2)\tilde{G}(k^2)}{\tilde{G}(p^2)}},
\label{result}
\end{eqnarray}
where $\tilde{G}$ is the Fourier image of some function 
$G_c^{-1},$ 
we can substitute this expression in  Eq. (\ref{basic}):
\begin{eqnarray}
&  \dis{\frac{\tilde{G}((p_2+k_2)^2)~  
\tilde{G}(k_2^2) }{\tilde{G}(p_2^2)} \times 
 \frac{\tilde{G}((p_2+k_2+k_1)^2)~\tilde{G}(k_1^2)}
{\tilde{G}((p_2+k_2)^2)}} = \no\\
& = \dis{\frac{\tilde{G}((p_2+k_2+k_1)^2)~  \tilde{G}((k_2+k_1)^2) }
  {\tilde{G}(p_2^2)}  } 
\times\dis{\frac{\tilde{G}(k_1^2)~\tilde{G}(k_2^2)}{\tilde{G}((k_1+k_2)^2)}. }
   \label{ex5}
\end{eqnarray}
This is an identity. That is, for the ansatz (\ref{result}), 
Eq. (\ref{basic}) is valid.
Now we will demonstrate that this anzatz is unique solution. 

In general, the function $\dis{\tilde{\G}(p,q,k)}$ 
is a function of three independent Lorentz invariants, since the moments 
$p$, $q$ and $k$ are not independent but related by conservation 
of the moments,~ $p + q + k = 0.$ We can choose $p^2$, $q^2$ and $k^2$
as those independent invariants, 
\begin{eqnarray*}
\dis{\tilde{\G}(p,q,k)} \equiv \dis{f(p^2,q^2,k^2)}. 
\end{eqnarray*}
Therefore, we can rewrite the basic equation (\ref{basic})
as 
\begin{eqnarray}
& \dis{f\le (p_2+k_2)^2,(p_2+k_2+k_1)^2,k_1^2\ri \times 
f\le p_2^2,(p_2+k_2)^2,k_2^2\ri}
\label{basic4}\\
& = \dis{f\le p_2^2,(p_2+k_2+k_1)^2,(k_1+k_2)^2\ri \times 
f\le (k_1+k_2)^2,k_1^2,k_2^2\ri}. \no
\end{eqnarray}

Let us introduce into the equation (\ref{basic4}) new independent 
variables, 
\begin{eqnarray}
& \dis{(p_2+k_2)^2 = x,~~~ (p_2+k_2+k_1)^2 = y,} \no\\
& \dis{k_1^2 = z,~~~ p_2^2 = u,}  \label{varD6}\\
& \dis{k_2^2 = v,~~~ (k_1+k_2)^2 = w.} \no
\end{eqnarray} 
The number of the independent variables is six, since in  (\ref{basic4})
we have only three independent Lorentz vectors $p_2, k_2, k_1.$
Using these vectors we can construct six Lorentz-invariant values above. 
In terms of these new independent variables the basic equation 
(\ref{basic4}) looks like 
\begin{eqnarray}
 \dis{f\le x,y,z\ri \times f\le u,x,v\ri} = 
\dis{f\le u,y,w\ri \times f\le w,z,v\ri}.
\label{basic5}
\end{eqnarray}

We consider equation (\ref{basic5}) as an equation for an analytical 
function of three variables in $R^3$ space. 
We observe that the r.h.s. of (\ref{basic5}) does not depend on $x$ 
for {\em any} values of $y,z,u,v.$ There is the unique solution to this 
- the dependence on $x$ must be factorized in the following way: 
\begin{eqnarray}
\dis{f\le x,y,z\ri = \frac{1}{\varphi(x)} F_1(y,z),~~~ 
f\le u,x,v\ri = \varphi(x) F_2(u,v),} \label{fact}
\end{eqnarray}
where ${\varphi}(x)$ is some function, and $F_1(y,z)$ and $F_2(u,v)$ are other 
functions. The rigorous prove of this statement is given below . The two 
equations in (\ref{fact}) imply 
\begin{eqnarray*}
\dis{f\le x,y,z\ri = \frac{\varphi(y)}{\varphi(x)}\times F(z)},
\end{eqnarray*}
where $F(z)$ is some function. By substituting this in Eq. (\ref{basic5}) 
we immediately infer that $F(z) = {\rm constant}\times\varphi(z).$ 
Rescaling $\varphi(z)$ by an appropriate constant, we obtain:
\begin{eqnarray}
\dis{f\le x,y,z\ri = \frac{\varphi(y)\varphi(z)}{\varphi(x)}}. \label{result2}
\end{eqnarray}
Let us give a rigorous proof that the factorization (\ref{fact}) 
of the $x$ dependence  
is the unique solution to  equation (\ref{basic5}). 
Denote $h \equiv \ln~f.$ Applying logarithm to (\ref{basic5}), we have 
\begin{eqnarray}
h(x,y,z) = - h(u,x,v) + ~{\rm terms~independent~on}~~ x. \label{1a}
\end{eqnarray}  
Applying $\dis{\frac{d^m}{dx^m}},$ $m = 1,2,...$ to (\ref{1a}), we 
obtain 
\begin{eqnarray*}
\frac{\pd^m h(x_1,y,z)}{\pd x_1^m}|_{x_1=x} = 
- \frac{\pd^m h(u,x_2,v)}{\pd x_2^m}|_{x_2=x}
\end{eqnarray*}
This means that the Taylor expansions in $x$ around the point $x=0$ 
for functions $h(x,y,z)$ and $h(u,x,v)$ are 
\begin{eqnarray}
& \dis{h(x,y,z) = h(0,y,z) - \tilde{\varphi}(x,y,z)}, \label{3a}\\
& \dis{h(u,x,v) = h(u,0,v) + \tilde{\varphi}(x,y,z)}, \label{3b}
\end{eqnarray}  
where 
\begin{eqnarray*}
\tilde{\varphi}(x,y,z) = - \sum_{n=1}^{\infty}\frac{1}{n!}x^n
\frac{\pd^n h(x_1,y,z)}{\pd x_1^n}|_{x_1=0}.
\end{eqnarray*}
Applying exponent to both the sides of (\ref{3a}) and (\ref{3b}),
we obtain  
\begin{eqnarray}
& \dis{f(x,y,z) = \frac{f(0,y,z)}{\varphi(x,y,z)}}, \label{4a}\\
& \dis{f(u,x,v) = f(u,0,v)\times \varphi(x,y,z)}, \label{4b}
\end{eqnarray}  
where $\varphi(x,y,z) = \exp{\tilde{\varphi}(x,y,z)}.$
 In (\ref{4b}) the l.h.s. is $y$- and $z$-independent. Hence, 
$\varphi(x,y,z)$ is also $y$- and $z$-independent: $\varphi(x,y,z) \equiv
\varphi(x).$ Thus, we can rewrite Eqs. (\ref{4a}) and (\ref{4b})
as Eq. (\ref{fact}), where  
\begin{eqnarray*}
F_1(y,z) \equiv f(0,y,z), ~~~ F_2(u,v) \equiv f(u,0,v).
\end{eqnarray*}
This proves (\ref{fact}) and thus (\ref{result2}). Thus, we can conclude 
from (\ref{result2}) that (\ref{result})
is the unique solution for $\tilde{\G}(p,q,k).$ To go back to the coordinate 
representation, we have to perform a Fourier transformation of (\ref{result}), 
\begin{eqnarray}
& \dis{\G (x,y,z) = \int~dpdqdk~\del(p+q+k)\tilde{\G}(p,q,k)
\exp(ipx +iqy + ikz)} \no\\
& = \dis{\int~dpdqdk~\del(p+q+k)\frac{\tilde{G}(q^2)
\tilde{G}(k^2)}{\tilde{G}(p^2)} 
 \exp(ipx +iqy + ikz)} \no\\
& =  \dis{\int~dx'dpdqdk~\exp(-i(p +q + k)x')\frac{\tilde{G}(q^2)
  \tilde{G}(k^2)}{\tilde{G}(p^2)}\exp(ipx +iqy + ikz)} \no\\
& = \dis{\int~dx'~G_{c}(x'-x)~G^{-1}_{c}(x'-y)
     ~G^{-1}_{c}(x'-z).} \label{result3}
\end{eqnarray}
By substituting this result in the second of equalities (\ref{equalities}), 
we can see that it is also satisfied by this solution. One can take the 
correct tree level normalization of $T^{abc}$  
\begin{eqnarray}
T^{abc} = \frac{i}{2}f^{bca} \label{assumption} 
\end{eqnarray}
and present the final result for the functional 
structure of $Lc^2$ proper correlator in the following form: 
\begin{eqnarray*}
& \dis{\int~dx~dy~dz~\G (x,y,z)T^{abc}
   L^{a}(x)c^{b}(y)c^{c}(z)} = \\
& = \dis{\int~dx'dxdydz~G_{c}(x'-x)~G^{-1}_{c}(x'-y)
~G^{-1}_{c}(x'-z)\frac{i}{2}f^{bca} L^{a}(x)c^{b}(y)c^{c}(z).} \no
\end{eqnarray*}

As we have mentioned above, the natural assumption (\ref{assumption}) 
about the group structure of the proper correlator $Lc^2$ 
has been done. However, we could avoid this assumption. Indeed, 
if all the group structures in (\ref{identitynew}) are independent, 
we obtain from there,  instead of (\ref{equalities}), three equations
\begin{eqnarray*}
& \dis{\int~dx~\G (y',x,z')\G (x,y,z) = \int~dx~\G (y',y,x)\G (x,z,z')} \\
& = \dis{\int~dx~ \G (y',x,z)\G (x,z',y)} = 0  
\end{eqnarray*} 
which are not true even at tree 
level as can be seen from Eq. (\ref{tree}). 
This means that at most two of the group structures must be 
independent to have a consistent solution. In such a case 
we come again to the necessity to solve Eq. (\ref{cond1})
that has unique solution (\ref{result3})   as we have demonstrated above. 
Substituting this solution in Eq.(\ref{identitynew})
we obtain Jacobi identities for $T^{abc}$ which means 
that they are structure constants. In detail, this procedure 
can be done as follows. We can substitute the result 
(\ref{result3}) in (\ref{A1})
\begin{eqnarray}
& \dis{\int~dx~dy~dz~\G (x,y,z)T^{abc}
   L^{a}(x)c^{b}(y)c^{c}(z)} \label{tree2}\\
& = \dis{\int~dx'dxdydz~G_{c}(x'-x)~G^{-1}_{c}(x'-y)
     ~G^{-1}_{c}(x'-z)T^{abc} L^{a}(x)c^{b}(y)c^{c}(z)} \no
\end{eqnarray}
and then redefine fields $L$ and $c$ 
\begin{eqnarray*}
& \dis{c^a(x)   = \int~d~x'~G_{c}(x-x')~\tilde{c}^a(x')} \\
& \dis{L^a(x)   = \int~d~x'~G^{-1}_{c}(x-x')~\tilde{L}^a(x')} \\
& \dis{\int~d~x'~G^{-1}_{c}(x-x')~G_{c}(x'-x'') = \del(x - x'').}  
\end{eqnarray*}
The second term in Slavnov--Taylor identity (\ref{STrMnon}) is covariant 
with respect to this change of variables, 
\begin{eqnarray}
\int~d~x~\frac{\del \G[L,c]}{\del c^a(x)}\frac{\del \G[L,c]}{\del L^a(x)} = 
\int~d~x~\frac{\del \G[L(\tilde{L}),c(\tilde{c})]}{\del \tilde{c}^a(x)}
\frac{\del \G[L(\tilde{L}),c(\tilde{c})]}{\del \tilde{L}^a(x)}, \label{oper}
\end{eqnarray}
as can be explicitly checked, but the expression (\ref{tree2}) 
takes the local form, 
\begin{eqnarray*}
\dis{\G  = \int~dx~T^{abc} \tilde{L}^{a}(x)\tilde{c}^{b}(x)\tilde{c}^{c}(x).} 
\end{eqnarray*}
By substituting this in the ST operator  (\ref{oper}) one concludes that
\begin{eqnarray}
T^{abc} = \frac{i}{2}f^{bca} \label{f}
\end{eqnarray}
solves it. The reason for this is that this $f^{abc}$ structure 
appears also at the level of the classical action 
\begin{eqnarray*}
2\Tr\int~dx~L(x)~c^2(x) = \frac{i}{2}f^{bca}L^{a}(x)c^{b}(x)c^{c}(x),
\end{eqnarray*}
and we already know that this structure satisfies the ST operator 
(\ref{oper}). Furthermore, there can be no other solution for 
$T^{abc},$ because (\ref{f}) is the only one 3-tensor in the adjoint 
representation of the gauge group that is antisymmetric in the last two
indices and satisfies Jacobi identities. Thus, the final result for 
the functional structure of $Lc^2$ proper correlator is 
\begin{eqnarray}
& \dis{\int~dx~dy~dz~\G (x,y,z)T^{abc}
   L^{a}(x)c^{b}(y)c^{c}(z)} = \label{structure2}\\
& = \dis{\int~dx'dxdydz~G_{c}(x'-x)~G^{-1}_{c}(x'-y)
~G^{-1}_{c}(x'-z)\frac{i}{2}f^{bca} L^{a}(x)c^{b}(y)c^{c}(z).}
\no
\end{eqnarray}

In conclusion of this section we present arguments that the form 
(\ref{structure2}) of the $Lcc$ correlator remains unchanged  
even if corrections from $LccA$  correlator are allowed to contribute 
to the $\sim Lc^3$ term in the ST equation, i. e. the first term in the 
ST identity (\ref{STrMnon}) contributes as well.  This results in 
corrections to Eq. (\ref{identitynew}). In such a case we 
can demonstrate that the basic equation (\ref{basic5}) will be 
modified to the following form:  
\begin{eqnarray}
& \dis{f\le x,y,z\ri \times f\le u,x,v\ri} -  
\dis{f\le u,y,w\ri \times f\le w,z,v\ri} \label{f2}\\ = 
& \dis{f_2\le u,z,v,y,x,w\ri - f_2\le u,v,z,y,u+z+y+v-x-w,w\ri.} \no
\end{eqnarray}
The new function $f_2$ of the variables (\ref{varD6}) parameterizes 
the contribution from the $LccA$ correlator. As one can see, there is 
a four-dimensional subspace of the six-dimensional space 
(\ref{varD6}) with coordinates $x,y,z,u,v,w$ which is intersection 
of two hyperplanes $x=u+z+y+v-x-w$
and $v=z$ where the contribution of $LccA$ in Eq. (\ref{f2}) disappears.
In this four-dimensional subspace Eq. (\ref{f2}) takes the same form 
as the basic equation (\ref{basic5}) takes in the six-dimensional space    
\begin{eqnarray}
& \dis{f\le \frac{u+2z+y-w}{2},y,z\ri \times f\le u,\frac{u+2z+y-w}{2},z\ri} 
\no\\
& -  \dis{f\le u,y,w\ri \times f\le w,z,z\ri}  = 0. \label{basic100}
\end{eqnarray}
Unfortunately, at present we do not have a clear proof that 
the factorization (\ref{result2}) is the only solution to this 
equation. However, there are several strong indications in favor of 
uniqueness of the factorization. Indeed, one of them is that if we reduce 
the subspace in consideration further to $u=y=z$ and $w=4\a z,$ 
where $\a$ is an arbitrary real parameter, we obtain 
\begin{eqnarray}
\dis{f\le 2(1-\a)z,z,z\ri \times f\le z,2(1-\a)z,z\ri} -  
\dis{f\le z,z,4\a z\ri \times f\le 4\a z,z,z\ri}  = 0. \no
\end{eqnarray} 
This suggests 
\begin{eqnarray*} 
\dis{\frac{d}{d\a}~f\le z,z,\a z\ri \times f\le \a z,z,z\ri = 0.}
\end{eqnarray*} 
As we have shown above, the factorization (\ref{result2})
is the only solution for such type of equations. 

Another indication in favor of the factorization (\ref{result2}) is that 
for the region of the four-dimensional subspace in consideration 
where $z$ is much larger than each of $u,$  $y,$  and  $w$ 
we have in the leading order of $u/z$ and  $y/z$ the equation 
\begin{eqnarray}
\dis{f\le z,y,z\ri \times f\le u,z,z\ri} -  
\dis{f\le u,y,w\ri \times f\le w,z,z\ri}  = 0, \no
\end{eqnarray}
that also requires the factorization (\ref{result2}) as the 
only solution since the information about $w$ disappears
on the l.h.s. 

As the third indication, we can decompose logarithm 
of Eq. (\ref{basic100}) in Taylor expansion in vicinity of 
any point in the four-dimensional subspace with coordinates 
$u,y,z,w.$ We then obtain, for the function $h = \ln f$ at the 
quadratic order of Taylor expansion, separability of the 
variables as the only solution. But the separability for $h$ 
means the factorization  for $f.$ Further, we have indications that
the separability must occur at any order of Taylor expansion.  

Thus, we have shown that there are at least three arguments in favor of the 
factorization (\ref{result2}) being the only solution also for the Eq. 
(\ref{basic100}), where this latter equation takes into account 
possible corrections from the $LccA$ correlator to the 
basic equation (\ref{basic5}).

\section{Solution to the correlator of $K_mA_mc$ type} \label{s4}

Starting from this point we can repeat the method that has been used 
in Ref. \cite{jhep} for deriving the solution to ST identities 
for supersymmetric theories. As it has been noted at the end of Introduction,
 the antighost equation (\ref{ghost}) restricts 
the dependence of $\G$ on the antighost field $b$ and on the external 
source $K_m$ to an arbitrary dependence on their combination 
\begin{eqnarray*}
\pd_m~b(x) + K_m(x). 
\end{eqnarray*}
We can present this dependence of the effective action on the external source 
$K_m$ in terms of a series  
\begin{eqnarray}
& \dis{\G = {\cal{F}}_0 + \sum_{n=1}\int d~x_1d~x_2\dots d~x_n 
{\cal{F}}^{m_1m_2\dots m_n}_n
\le x_1, x_2,\dots, x_n\ri \le\pd_{m_1}~b(x_1) + K_{m_1}(x_1)\ri \times} \no\\
& \dis{\times \le\pd_{m_2}~b(x_2) + K_{m_2}(x_2)\ri\dots 
\le\pd_{m_n}~b(x_n) + K_{m_n}(x_n)\ri,} \label{ser} 
\end{eqnarray} 
where we assume contractions in spacetime indices $m_j.$ The coefficient 
functions of this expansion are in their turn 
functionals of other effective fields (\ref{defphi}), 
\begin{eqnarray*}
{\cal{F}}^{m_1m_2\dots m_n}_n = 
{\cal{F}}^{m_1m_2\dots m_n}_n\left[A_m,c,L\right],    
\end{eqnarray*}
whose coefficient functions for example in case $L =0$ are  
ghost-antighost-vector correlators. ${\cal{F}}_0$ is a 
$K_m$-in\-de\-pen\-dent part of the effective action. The 
spacetime indices $m_j$ of ${\cal{F}}_n$ will be omitted 
everywhere below since they are not important in the present 
analysis. 

Our purpose is to restrict the expansion (\ref{ser}) by using 
the ST identities (\ref{STrMnon}). Let us consider for the moment 
the terms of (\ref{ser}) without the field $L.$   
The noninvariance of these terms with respect to the ST identities
(\ref{STrMnon}) must be compensated by the first term (\ref{firstLc^2}) 
of the series (\ref{dec}) or possible interactions of this term with 
physical effective fields because
$\dis{\frac{\del\G}{\del L(x)}}$ of such terms only  
has no $L.$ The total degree of the ghost fields $c$  
in ${\cal{F}}_n$ must be equal to $n$ since each proper graph contains 
equal number of ghost and antighost fields among its external legs.  

Let us consider terms in the effective action whose variations 
are cancelled by variations of the ghost field caused by the first 
term (\ref{firstLc^2}) of the series (\ref{dec}). To start we consider 
the ${\cal{F}}_1(x_1)$ coefficient function in the 
expansion (\ref{ser}). The corresponding term of the lowest order in fields 
in (\ref{ser}) is   
\begin{eqnarray}
& \dis{\int~d~x~d~x'~2i~\Tr\left[\le\pd_m~b(x) + K_m(x)\ri 
~\pd_m~G(x-x')~c(x')\right]}, \label{dos} 
\end{eqnarray}
where $-i\Box~G(x-x')$ is a 2-point ghost-antighost 
proper correlator. It is an Hermitian kernel of the above integral,      
\begin{eqnarray*}
G^{\dg} = G.
\end{eqnarray*} 

We can make any change of variables in the effective action $\G.$ 
Let us make the following change of variables:
\begin{eqnarray}
\begin{array}{cc}
 \dis{A_m(x) = \int~d~x'~G_{A}(x-x')~\tilde{A}_m(x')},   
& \dis{K_m(x) = \int~d~x'G^{-1}_{A}(x-x')~\tilde{K}_m(x')}, \\
 \dis{c(x)   = \int~d~x'~G_{c}(x-x')~\tilde{c}(x')}, 
& \dis{L(x)   = \int~d~x'~G^{-1}_{c}(x-x')~\tilde{L}(x')}, \\
  \dis{b(x) = \int~d~x'~G^{-1}_{A}(x-x')~\tilde{b}(x')}.
& ~~~  \\
\end{array} \label{change2}
\end{eqnarray}
Here $G_{X}(x-x')$ are some dressing functions,\footnote{The formula 
(\ref{bilocal2}) does not mean that both the functions 
$G^{-1}_{X}(x-x')$ and $G_{X}(x'-x'')$ are $\del$-functions. It means only 
that the product of their Fourier transforms is equal to 1.} 
\begin{eqnarray}
\int~d~x'~G^{-1}_{X}(x-x')~G_{X}(x'-x'') = \del(x - x''), \label{bilocal2} 
\end{eqnarray}
In terms of new variables the effective action $$\tilde{\G}[\tilde{\varphi},
\tilde{K}_m, \tilde{L}] 
= \G[\varphi(\tilde{\varphi}), K_m(\tilde{K}_m), L(\tilde{L})]$$ 
must satisfy the identity 
\begin{eqnarray}
& \Tr\left[\dis{\int~d~x~\frac{\del \tilde{\G}}{\del \tilde{A}_m(x)}
   \frac{\del \tilde{\G}}{\del \tilde{K}_m(x)}
  + \int~d~x~\frac{\del \tilde{\G}}{\del \tilde{c}(x)}
\frac{\del \tilde{\G}}{\del 
   \tilde{L}(x)}} \right. \label{STrMnoninnew} \\
& \dis{\left. 
  - \int~d~x~d~x'~d~x''~\frac{\del \tilde{\G}}{\del \tilde{b}(x')}
    G_{A}(x-x')\le\frac{1}{\a}~\pd_m~\tilde{A}_m(x'')
G_{A}(x-x'')\ri \right] = 0} \no 
\end{eqnarray} 
which is the identity (\ref{STrMnon}) re-expressed in terms of the new 
variables according to  (\ref{change2}). As one can see the ST operator 
is covariant with respect to this change of variables except for the gauge 
fixing term, which remains unaffected by quantum corrections anyway as
mentioned earlier.

One can make the  change of variables (\ref{change2}) in the 
integral (\ref{dos}).
\begin{eqnarray}
& \dis{\int~d~x~d~x'~d~x''~d~x'''~2i~\Tr\left[\le\pd_m~\tilde{b}(x'')
 + \tilde{K}_m(x'')\ri ~G^{-1}_{A}(x''-x)~G(x-x')\right.}~\times~\no\\
& \dis{\left.\times~ G_{c}(x'-x''')~\pd_m~\tilde{c}(x''')\right]}. 
\label{before} 
\end{eqnarray}
While the dressing function $G_{c}(x-x')$ has been defined
through the solution (\ref{firstLc^2}) to the operator
(\ref{STrMnonsmall}), the dressing function $G_{A}(x - x')$ has
not been defined yet. We define it from the requirement 
\begin{eqnarray*}
& \dis{\int~d~x~d~x' 
G^{-1}_{A}(x''-x)~G(x-x')~G_{c}(x'-x''') = \del(x''-x''').} 
\end{eqnarray*}
In such case the term (\ref{before}) after the change of variables
(\ref{change2}) simplifies to  
\begin{eqnarray}
& \dis{\int~d~x~2i~\Tr\left[\le\pd_m~\tilde{b}(x) + \tilde{K}_m(x)\ri~
\pd_m~\tilde{c}(x)\right]}. \label{aux3}  
\end{eqnarray}

The first term in ST identities (\ref{STrMnoninnew})
can also be expanded in terms of $\pd_m~\tilde{b}(x) + \tilde{K}_m(x),$  
\begin{eqnarray}
& \dis{\int~d~x~\frac{\del \tilde{\G}}{\del \tilde{A}_m(x)}
   \frac{\del \tilde{\G}}{\del \tilde{K}_m(x)}   = {\cal{M}}_0
+  \sum_{n=1}\int d~x_1d~x_2\dots d~x_n {\cal{M}}^{m_1m_2\dots m_n}_n
\le x_1, x_2,\dots, x_n\ri} \no\\
& \dis{\times\le\pd_{m_1}~\tilde{b}(x_1) + \tilde{K}_{m_1}(x_1)\ri
\le\pd_{m_2}~\tilde{b}(x_2) + \tilde{K}_{m_2}(x_2)\ri\dots\times} \no\\
& \dis{\times\dots\le\pd_{m_n}~\tilde{b}(x_n) + \tilde{K}_{m_n}(x_n)\ri}, 
\label{dos2}  
\end{eqnarray}
where we assume contractions in spacetime indices $m_j.$ Again, the 
spacetime indices $m_j$ of ${\cal{M}}_n$ will be omitted 
everywhere below since they are not important in the present 
analysis. ${\cal{M}}_0$ is the $\Kt_m$-independent part of 
(\ref{dos2}).  We can consider that the l.h.s. of 
(\ref{dos2}) is the result of an infinitesimal transformation in 
$\tilde{\G},$ in which instead of $\tilde{A}_m(x)$ we have substituted  
\begin{eqnarray}
\tilde{A}_m(x) \rar \tilde{A}_m(x) 
+ \frac{\del \tilde{\G}}{\del \Kt_m(x)}. \label{subs}
\end{eqnarray}
In other words, one can consider the result of such a substitution 
as the difference  
\begin{eqnarray*}
\G\left[ \Kt_m,\tilde{A}_m(x)+\frac{\del\G}{\del \Kt_m(x)}\right] 
- \G\left[ \Kt_m,\tilde{A}_m(x)\right] 
\end{eqnarray*}
to linear order in $\dis{\frac{\del\G}{\del \Kt_m(x)}}.$ Eq. (\ref{aux3})
implies that the ``gauge'' transformation (\ref{subs}) can be rewritten as 
\begin{eqnarray*}
\del\tilde{A}_m(x) \sim i\pd_m\tilde{c}(x) + {\rm higher~~terms}. 
\end{eqnarray*}
The sum of the part quadratic in $\tilde{A}$  of ${\cal{F}}_0$ and 
the ${\cal{F}}_1$-type term (\ref{aux3}) contributes to ${\cal{M}}_0$
by yielding a term $\sim \tilde{A}\tilde{c}.$ However, ${\cal{M}}_0$
must be equal to zero\footnote{In principle, another term $\sim \tilde{A}\tilde{c}$
can appear in the third term there, coming from the 
$\sim \tilde{b}\pd_m\tilde{\nabla}_m\tilde{c}$ part  of $\tilde{\G}.$ 
However, the third term in (\ref{STrMnoninnew}) [and in(\ref{STrMnon})]
is only responsible  for the absence of corrections to the gauge fixing 
term in $\G,$ as we have already noted at the end of Section \ref{s2}.}. 
Hence, the part quadratic in $\tilde{A}$  of ${\cal{F}}_0$ must be invariant,
at quadratic order, under the aforementioned ``gauge'' transformation,
implying the form
\begin{eqnarray}
\dis{~-\int d~x~Z_{g^2}\frac{1}{2g^2}~\Tr\le\pd_m\tilde{A}_n(x) - 
\pd_n\tilde{A}_m(x)\ri {\cal O}
\le\pd_m\tilde{A}_n(x) - \pd_n\tilde{A}_m(x)\ri},  \label{aux2}
\end{eqnarray}
where $Z_{g^2}$ is a number that depends on couplings and a regularization 
parameter of the theory, and ${\cal O}$ is some differential  
operator. Later we will see how the ST identities put restrictions on such 
an operator. 

Having fixed the form of the quadratic term (\ref{aux3}) in ${\cal{F}}_1,$ we consider 
the vertex of next order in fields in ${\cal{F}}_1,$ which looks like 
$\sim \le\pd_m\tilde{b} +  \Kt_m\ri\tilde{A}_m\tilde{c}.$ We will show 
now that the structure of the vertex 
$\sim \le\pd_m\tilde{b} +  \Kt_m\ri\tilde{A}_m\tilde{c}$ 
is fixed completely by the quadratic term (\ref{aux3}) and by the 
term (\ref{firstLc^2}). According to the 
Slavnov--Taylor identity (\ref{STrMnoninnew}), the contribution 
of $\sim \le\pd_m\tilde{b} +  \Kt_m\ri\tilde{A}_m\tilde{c}$ of 
the  ${\cal{F}}_1$ part of the effective action into ${\cal{M}}_1$
caused by the 
quadratic term (\ref{aux3}) due to the substitution (\ref{subs}) must 
be cancelled by the variation of the ghost field  caused in 
(\ref{aux3}) by the first term (\ref{firstLc^2}) of the series 
(\ref{dec}) due to substitution (\ref{aux4}). According to our 
conjecture, the term  
(\ref{firstLc^2}) has the form 
\begin{eqnarray*}
\dis{2\Tr \int~dx~\tilde{L}(x)\tilde{c}^2(x).} 
\end{eqnarray*}
Indeed, the only contribution to ${\cal{M}}_1$ of the order 
of $\sim \le\pd_m\tilde{b} +  \Kt_m\ri\pd_m\tilde{c}^2$ 
in ${\cal{M}}_1$ 
comes from this $\sim \le\pd_m\tilde{b} +  \Kt_m\ri\tilde{A}_m\tilde{c}$
term in ${\cal{F}}_1$:  
\begin{eqnarray*}
\dis{\int~d~x~\frac{\del {\tilde{\G}}|_{{\cal{F}}_1}} {\del \tilde{A}_m(x)}
 \frac{\del{\tilde{\G}}|_ {{\cal{F}}_1}}{\del \tilde{K}_m(x)} \sim 
 \left[\le\pd_m\tilde{b} +  \Kt_m\ri\tilde{c}\right]\pd_m\tilde{c} \sim 
 \le\pd_m\tilde{b} +  \Kt_m\ri\pd_m\tilde{c}^2}, 
\end{eqnarray*}
where $ {\tilde{\G}}|_{{\cal{F}}_1}$ is the ${\cal{F}}_1$ part of the
effective action. 
One could think at first that the ${\cal{F}}_0-$ and ${\cal{F}}_2-$ type 
terms $ {\tilde{\G}}|_{{\cal{F}}_0}$, $ {\tilde{\G}}|_{{\cal{F}}_2}$
of (\ref{dec}) might also contribute to the term of order  
$\sim \le\pd_m\tilde{b} +  \Kt_m\ri\pd_m\tilde{c}^2$ in ${\cal{M}}_1$
via 
\begin{eqnarray*}
\dis{\int~d~x~\frac{\del {\tilde{\G}}|_{{\cal{F}}_0}} {\del \tilde{A}_m(x)}
\frac{\del{\tilde{\G}}|_ {{\cal{F}}_2}}{\del \tilde{K}_m(x)}}
\end{eqnarray*}
because 
\begin{eqnarray*}
\dis{\frac{\del{\tilde{\G}}|_ {{\cal{F}}_2}}{\del \tilde{K}_m(x)} \sim
\le\pd_m\tilde{b} +  \Kt_m\ri{\cal{F}}_2[A_m,c]}.
\end{eqnarray*}
However, $\dis{\frac{\del {\tilde{\G}}|_{{\cal{F}}_0}} {\del \tilde{A}_m(x)}}$
starts with terms linear in $\tilde{A}_m(x).$ Thus, the ${\cal{F}}_2$
part of the effective action does not contribute to the term of the order of
$\sim \le\pd_m\tilde{b} +  \Kt_m\ri\pd_m\tilde{c}^2$ in  ${\cal{M}}_1,$
only the ${\cal{F}}_1$ part of the effective action does. Hence, the term 
of  order 
$\sim \le\pd_m\tilde{b} +  \Kt_m\ri\tilde{A}_m\tilde{c}$   in  ${\cal{F}}_1$ 
is the term of the same order that is contained in  
$\Kt_m(x)\tilde{\nabla}_m~\tilde{c}(x)$ because  only in this case 
the terms $\sim \le\pd_m\tilde{b} +  \Kt_m\ri\pd_m\tilde{c}^2$ in ${\cal{M}}_1$
will be cancelled by the second term in ST identities (\ref{STrMnoninnew}),
which will result in 
\begin{eqnarray*}
\dis{\int~d~x~2i~\Tr\left[\le\pd_m~\tilde{b}(x) + \tilde{K}_m(x)\ri~
\pd_m~\tilde{c}^2(x)\right]}   
\end{eqnarray*}
due to substitution (\ref{aux4}). Thus, the term of  
lowest order in fields in ${\cal{F}}_1$ is 
\begin{eqnarray}
2i~\Tr\left[\le\pd_m~\tilde{b}(x) + \tilde{K}_m(x)\ri\tilde{\nabla}_m
~\tilde{c}(x)\right], 
~~~ \tilde{\nabla}_m = \pd_m + i \tilde{A}_m.  \label{aux}
\end{eqnarray} 
All the terms in ${\cal{F}}_0$ of higher orders in $\tilde{A}_m(x)$ 
are fixed by themselves in an iterative way due to  
the requirement that ${\cal{F}}_0$ must be invariant 
with respect to the substitution (\ref{subs}). Taking into  
account (\ref{aux}), we see that the first invariant term is   
\begin{eqnarray*}
\dis{~-\int d~x~Z_{g^2}\frac{1}{2g^2}~\Tr\tilde{F}_{mn}(x)\tilde{F}_{mn}(x)},
\end{eqnarray*}
where $\tilde{F}_{mn}(x)$ is Yang--Mills tensor of $\tilde{A}_m(x).$ 
That is, the physical part of the effective action can be restored 
from the requirement of its invariance with respect to the gauge invariance 
in terms of the gauge field dressed by the dressing function. Here we 
see that the differential operators ${\cal O}$ in Eq. (\ref{aux2}) between two 
Yang--Mills tensors must be covariant derivatives. For example, the following 
term is allowed, 
\begin{eqnarray}
\dis{~\int d~x~f_2~\frac{1}{\Lambda^2}\Tr\tilde{F}_{mn}(x)
\tilde{\nabla}^2\tilde{F}_{mn}(x)}, \label{sup1}
\end{eqnarray}
where $f_2$ is another number that depends on couplings, and $\Lambda$ is 
a regularization parameter of the theory. Starting from the fourth degree 
of $\tilde{A}_m(x) $ higher order gauge invariant contributions like  
\begin{eqnarray}
\int~d~x~f_3~\frac{1}{\Lambda^4}\Tr~\tilde{F}_{mn}(x)\tilde{F}_{mn}(x)
\tilde{F}_{kl}(x)\tilde{F}_{kl}(x) \label{sup2}  
\end{eqnarray}
into ${\cal{F}}_0$ are allowed. Here $f_3$ is another number that depends 
on couplings.

\section{Further steps for higher correlators in $K_m$ and $L$} \label{s5}

We consider now the coefficient functions ${\cal{F}}_n$ with $n > 1$
in (\ref{ser}) for $L = 0.$ There are two possibilities here. 
The first possibility is that these terms of higher degrees in $\Kt$
do not respect the gauge invariance of the physical part of (\ref{ser})
created by the ${\cal{F}}_1$ term. In such a case ${\cal{F}}_2$ contributes 
to ${\cal{M}}_1$ but we do not have anything that can compensate 
this contribution by ghost transformations induced by the second term in the 
ST identities (\ref{STrMnoninnew}). Hence,  ${\cal{F}}_2 = 0.$ If we 
consider ${\cal{F}}_3,$ it 
contributes to  ${\cal{M}}_2$ and, in general, could be compensated 
by ghost transformations in ${\cal{F}}_2$. But ${\cal{F}}_2$ is zero,
hence, ${\cal{F}}_3$ is also zero. We can repeat the former argument 
for all higher numbers $n$ of ${\cal{F}}_n.$ All coefficient functions 
${\cal{F}}_n$ with $n > 1$ are equal to zero in the first possibility.
The second possibility is that the terms of higher degrees in $\Kt$
respect the gauge invariance of the physical part of (\ref{ser}).
In such a case ${\cal{F}}_n$ with $n > 1$ does not contribute 
to ${\cal{M}}_n$ for any $n.$ In supersymmetric theories 
this possibility does not exist \cite{jhep} due to chiral nature 
of the ghost superfields. However, in the nonsupersymmetric case
one can invent, for example, ${\cal{F}}_2$ constructions such as the 
following one
\begin{eqnarray}
\int~d~x~\Tr\left[\le \tilde{c}(x)\tilde{\nabla}_m
\le\pd_m~\tilde{b}(x) + \tilde{K}_m(x)\ri\ri
\le \tilde{c}(x)\tilde{\nabla}_m\le\pd_m~\tilde{b}(x) + 
\tilde{K}_m(x)\ri\ri\right]. 
\label{ex}
\end{eqnarray}
Such a term gives zero contribution to ${\cal{M}}_1,$ since 
its variation with respect to $\Kt$ is proportional to 
$\tilde{\nabla}_m({\rm scalar~function})$ and its 
contribution to ${\cal{M}}_2$ can be cancelled by the transformation of 
the ghost field in ${\cal{F}}_2$ if the coefficient before (\ref{ex})
has been fixed in an appropriate way. This can be proved in the same way 
(\ref{BRSTtrick}) which has been used to derive the BRST 
transformation in Section \ref{s2}. 

We have considered the terms in the effective action whose variations 
are cancelled by variations of the ghost field caused by the first 
term (\ref{firstLc^2}) of the series (\ref{dec}). In general, some 
sophisticated interactions of the term (\ref{firstLc^2}) with 
physical fields can be introduced. However, again we can state 
that the higher order terms must respect the already established invariance 
with respect to the Slavnov--Taylor operator for the terms of lowest 
degrees in fields. In our case for example we can write for 
interactions of the term (\ref{firstLc^2}) with physical fields
by using the following substitution 
\begin{eqnarray*}
\dis{\tilde{L}\tilde{c}^2 \rar \tilde{L}\tilde{c}^2
\le 1 + f_4~\frac{1}{\Lambda^4}\Tr\tilde{F}_{mn}(x)\tilde{F}_{mn}(x)\ri},
\end{eqnarray*}
and then  making a substitution in (\ref{aux}):
\begin{eqnarray}
\dis{\tilde{c} \rar \tilde{c}
\le 1 + f_4~\frac{1}{\Lambda^4}\Tr\tilde{F}_{mn}(x)\tilde{F}_{mn}(x)\ri}. 
\label{ex2}
\end{eqnarray}
However, these terms cannot change the structure 
of the physical part of the effective action since it is already 
determined by the terms of the first order in the auxiliary field 
$\Kt_m.$ 
  
One can consider possible terms with higher degrees of $L.$
For example, the sum of (\ref{firstLc^2}) and 
\begin{eqnarray}
\int~dx~\sum_{a;~b_1,b_2,\dots,b_{4k}}
\le \tilde{L}^{a}(x)\tilde{L}^{a}(x) \ri^{k} 
\tilde{c}^{b_1}(x)\dots\tilde{c}^{b_{4k}}(x)\e_{b_1 b_2\dots b_{4k}} 
\label{ex3}
\end{eqnarray}
satisfies the identity (\ref{STrMnonsmall}) if $4k$ is the rank
of the gauge group. If these terms exist it is also necessary 
to consider the dependence of ${\cal{F}}_n$ on the auxiliary field $L,$ 
since the substitution due to the second term in the ST identities 
would produce these terms. However, at the end we put all the 
auxiliary fields equal to zero, and therefore all 
the terms with higher degrees of $\tilde{L}$  do not have 
any importance. In comparison, the situation with 
the $\Kt_m$ field is different. Indeed, terms with zero $\Kt_m$ are 
still important since they are responsible for higher 
degrees of ghost-antighost correlators which may have applications in 
some models.

\section{Conjecture for the physical part of the action} \label{s6}

Taking into account the structure (\ref{aux}) of the term linear
in $\Kt_m,$ one can come to a natural conjecture about the 
form of the part of the effective action that depends only on 
the gauge effective field  $A_m.$ Namely, due to the ST identity 
(\ref{STrMnoninnew}) in terms of the dressed fields, the structure of 
the effective action is  
\begin{eqnarray}
& \dis{\G[A_m,b,c] = 
\int d~x~\left[-\frac{1}{2g^2}~Z_{g^2}\Tr\le\tilde{F}_{mn}(x)~{\cal G}\le 
\frac{\tilde{\nabla}^2}{\Lambda^2}\ri\tilde{F}_{mn}(x)\ri  
 \right.} \label{G}\\
& \dis{\left. - ~\Tr\le\frac{1}{\a}\left[\pd_m A_m(x)\right]^2 \ri
 - 2i~\Tr~\le\tilde{b}(x)\pd_m~\tilde{\nabla}_m~\tilde{c}(x)\ri\right] +
\rm{irrelevant}\ \   \rm{part}} \no,
\end{eqnarray}   
where all auxiliary fields $K$ and $L$ are set equal to zero. 
It is necessary to make three comments here: 
\begin{itemize}
\item The function ${\cal G}$ is a series in terms of covariant 
    derivative with dressed gauge connection. The part of this series
     without 
    gauge connection ${\cal G}\le \frac{\pd^2}{\Lambda^2}\ri$
    has logarithmic asymptotic in the momentum space at high momentum,      
    ${\cal G}\le \frac{-p^2}{\Lambda^2}\ri \sim \ln\le -\frac{p^2}
    {\Lambda^2}\ri,$
    while at low momentum it may be represented,e.g., by powers of 
    $p^2/\Lambda_{QCD}^2$ with $\Lambda_{QCD} \sim 0.1$  GeV \cite{second}.
\item The physical part of the action is gauge invariant in terms of 
      the dressed field $\tilde{A}_m(x).$
\item We do not write in the physical part terms like (\ref{sup2}) 
    since finally we are going to take the regularization mass 
    $\Lambda$ to infinity. Terms like  (\ref{sup2}), 
    (\ref{ex}), (\ref{ex2}) are called irrelevant in (\ref{G}).
\end{itemize}
      
\section{Regularization and renormalization} \label{s7}
 
In a general nonsupersymmetric four-dimensional gauge theory
which is regularized in a way that preserves gauge (and BRST)
symmetry, the dressing functions are of the following form: 
\begin{eqnarray}  
& \dis{G^{-1}_{X}(x-x') = z_X \del(x-x') + 
\frac{C_1(\Lambda^2,\mu^2)}
{\mu^2}\le\Box - \mu^2\ri\del(x-x')} \no\\
& + \dis{\frac{C_2(\Lambda^2,\mu^2)}{\le\mu^2\ri^2}\le\Box - 
\mu^2\ri^2 \del(x-x') + ... } \label{exp}
\end{eqnarray}
This representation means that we have expanded the Fourier transformed
dressing function $\tilde{G}^{-1}_{X}(p^2) = 1/\tilde{G}_{X}(p^2),$ 
$X=A,c,$ in the vicinity of the 
point $p^2 = - \mu^2.$ Here $z_X$ is a constant that goes to infinity 
if the regularization is removed, and $C_1,$ $C_2$ are finite 
constants.\footnote{$G^{-1}_{X}(x-x') = 
(2\pi)^{-4}\int~dp\exp{[-ip(x-x')]}(1/\tilde{G}_{X}(p^2))$, i.e.
$G^{-1}_{X}(x-x') \ne 0$ for $x-x' \ne 0$ in general, although the expansion 
(\ref{exp}) might suggest otherwise.} For instance, $z_A$ is 
a renormalization constant of the gauge field. To renormalize the 
theory we have to introduce counterterms into the classical 
action (\ref{action}) \cite{BoSh}. This is equivalent to the change 
of the field in the classical action (\ref{action}). For example, 
in the case of the pure gauge theory, to remove divergences from 
$G^{-1}_{A}(x-x')$ we have to make the following redefinition 
of the gauge field in the classical action 
\begin{eqnarray} 
\dis{A_m^{\rm bare} \rar \frac{A_m^{\rm phys}}{z_A}}. \label{counter}
\end{eqnarray}
The motivation for terminology ``bare'' and ``physical'' for fields in the 
path integral is that introducing counterterms into the classical action 
(\ref{action}) by the rescaling (\ref{counter}) of fields and couplings 
will result in the effective action without divergences (renormalized 
effective action). We can show that by such a redefinition we can make 
the dressing function $G^{-1}_{A}$ finite. Indeed, 
if we represent the term with the source of the gauge field 
in the path integral (\ref{pathRMnon}) as 
\begin{eqnarray*} 
\dis{J_m~A_m = \le J_m~z_A\ri\frac{A_m}{z_A}},
\end{eqnarray*}
then the path integral for the theory with counterterms 
(\ref{counter}) can be transformed to the form (\ref{pathRMnon})
by the substitution of variables of the integral  
$\dis{A_m = A_m'z_A.}$ 
It means that all the previous construction 
can be reproduced without any change but taking into account the 
redefinition $\dis{J_m \rar J_m~z_A.}$
In turn such a redefinition, according to definitions (\ref{defphi}),
means nothing else but that the effective fields are also redefined as  
in (\ref{counter}), which is equivalent to the redefinition of 
the dressing function 
\begin{eqnarray}
G^{-1}_{A}(x-x') \rar \frac{1}{z_A}G^{-1}_{A}(x-x'). 
\label{red}
\end{eqnarray}
One can consider Eq. (\ref{red}) in  momentum space, 
\begin{eqnarray} 
& \dis{\frac{1}{z_A}~\frac{1}{\tilde{G}_{A}(p^2)} = 
\frac{\tilde{G}_{A}(-\mu^2,\Lambda^2)}
{\tilde{G}_{A}(p^2,\Lambda^2)} = 
\le 1 +\alpha g^2  + \beta \le g^2\ri^2  +  \gamma \le g^2\ri^3 + \dots \ri} 
  \times \no\\ 
& \times\dis{\le 1 + \tilde{G}_{1}(p^2) g^2 + 
 \tilde{G}_{2}(p^2)\le g^2\ri^2 +\dots \ri}  \label{R}\\
& \dis{= 1 + \le \alpha + \tilde{G}_{1}(p^2) \ri g^2 + 
  \le \beta + \alpha\tilde{G}_{1}(p^2) + \tilde{G}_{2}(p^2) \ri
  \le g^2\ri^2 + \dots} \no
\end{eqnarray}
where we have presented both factors on the l.h.s. as a series in terms 
of the coupling constant. In this expansion $g^2$ is the physical coupling  
that stays in the classical action according to the counterterm approach
\cite{BoSh}. All these dressing functions parametrize our result (\ref{G})
for the effective action, that is, they parametrize the irreducible vertices
that contain divergences. Divergences from the dressing functions 
must be removed. We can remove the 
divergences in each order in coupling constant by choosing the 
divergent coefficients $\alpha,$ $\beta,$ $\gamma$ in $1/z_A$ in an 
appropriate way, because each coefficient  $\tilde{G}_{n}(p^2)$ of the 
decomposition   $\tilde{G}^{-1}_{A}(p^2)$ in terms of the coupling 
constant is in its turn a series in terms of $p^2$ with only the zero order 
in $p^2$ terms being divergent. This is due to the fact that 
\begin{eqnarray*} 
\dis{\lim_{\Lambda \rar \infty} 
\frac{\tilde{G}_{A}(p^2,\Lambda^2)}
{\tilde{G}_{A}(-\mu^2,\Lambda^2)}}
\end{eqnarray*}
is finite. As to divergent coefficients before the relevant operators, they 
will be compensated by counterterms from the bare couplings
\footnote{Even if the renormalization
(\ref{R}) has been done and the dressing functions are finite, the theory 
still has divergences in the coefficients of the relevant operators.
These divergences are absorbed by the bare couplings.}. 

Till this moment we did not specify which regularization is used. 
The regularization by higher derivatives (HDR) is the most convenient from the 
point of view of the theoretical analysis \cite{SF}. It provides 
strong suppression of ultraviolet divergences by introducing  
additional terms with higher degrees of covariant derivatives acting on 
Yang--Mills tensor into the classical action (\ref{action}), which are 
suppressed by appropriate degrees of the regularization scale 
$\Lambda.$  In addition to this it 
is necessary to introduce a modification of the Pauli--Villars
regularization to guarantee the convergence of the one loop 
diagrams \cite{SF}. To regularize the fermion cycles, the usual 
Pauli--Villars regularization can be used.\footnote{A somewhat different 
regularization approach is applied in Ref. \cite{second} where explicit 
QCD one loop dressing functions are obtained.} 

Thus in case of four-dimensional QCD without quarks the classical 
action (\ref{action}) is 
\begin{eqnarray*}
& \dis{S_{QCD}[A,b,c] = \int~d~x~~
\left[-\frac{1}{2g^2}~\Tr\left[F_{mn}(A(x))F_{mn}(A(x))\right]\right.} \\ 
& \dis{\left. - ~\Tr\le\frac{1}{\a}\left[\pd_m A_m(x)\right]^2 \ri   
   - 2~\Tr~\le~i~b(x)\pd_m~\nabla_m(A)~c(x)\ri \right]}. 
\end{eqnarray*} 
In the counterterm technique \cite{BoSh} the coupling constant 
here is the physical coupling constant. The classical action with the 
counterterms is 
\begin{eqnarray*}
& \dis{S_{QCD}[A,b,c] = \int~d~x~~
\left[-\frac{1}{Z_{g^2}}\frac{1}{2g^2}~\Tr\left[F_{mn}\le\frac{A}{z_A}(x)\ri
F_{mn}\le\frac{A}{z_A}(x)\ri\right]\right.} \\
& \dis{\left. - ~\Tr\le\frac{(z_A)^2}{\a}\left[\pd_m \frac{A_m}{z_A}(x)\right]^2 \ri   
   - 2~\Tr~\le~i~z_Ab(x)\pd_m~\nabla_m\le\frac{A}{z_A}\ri~\frac{c}{z_c}(x)\ri \right]}, 
\end{eqnarray*} 
where fields are ``physical'' in the sense that this classical 
action together with counterterms results in the effective action 
in which divergences are removed. Thus, we come to 
the conclusion that the {\em renormalized} effective action takes 
the form:
\begin{eqnarray}
& \dis{\G_{QCD}[A_m,b,c] = 
\int d~x~\left[-\frac{1}{2g^2}~\Tr\le\tilde{F}_{mn}(x){\cal G}_2\le 
\frac{\tilde{\nabla}^2}{\mu^2}\ri\tilde{F}_{mn}(x)\ri  
 \right.} \label{QCD}\\
& \dis{\left. - ~\Tr\le\frac{1}{\a}\left[\pd_m A_m(x)\right]^2 \ri
 - 2i~\Tr~\le\tilde{b}(x)\pd_m~\tilde{\nabla}_m~\tilde{c}(x)\ri\right]} \no,
\end{eqnarray}
where all auxiliary fields   
$K$ and $L$ are set equal to zero. Here the function  ${\cal G}_2$ is 
defined as 
\begin{eqnarray}
{\cal G}_2\le \frac{\tilde{\nabla}^2}{\mu^2}\ri \equiv 
\lim_{\Lambda \rar \infty} 
{\cal G}\le \frac{\tilde{\nabla}^2}{\Lambda^2}\ri/
{\cal G}\le \frac{\mu^2}{\Lambda^2}\ri.
\end{eqnarray}

\section{Summary} \label{s8}
  
In this work we proposed a solution to the Slavnov--Taylor identities 
for the effective action of nonsupersymmetric non-Abelian gauge theory
without matter. The solution is expressed in terms of gauge $A_m$ and 
(anti)ghost effective fields $(c,b)$ convoluted with unspecified 
dressing functions: 
\begin{eqnarray*} 
& \dis{\tilde{A}_m(x) = \int~d~x'~G^{-1}_{A}(x-x')~A_m(x')} \\   
& \dis{\tilde{c}(x)   = \int~d~x'~G^{-1}_{c}(x-x')~c(x')}, \\
& \dis{\tilde{b}(x) = \int~d~x'~G_{A}(x-x')~b(x')}.
\end{eqnarray*} 
Further, the solution is 
invariant under the gauge (BRST) transformation of the convoluted fields. 
We gave arguments which show that, under a specific plausible assumption,
the terms of the effective action containing (anti)ghost fields must have 
the same form as those in the classical action, but under the substitution 
$X \rar \tilde{X}$ $\le X = c,b,A_m\ri.$ Further, we conjectured 
a rather general form of the terms of the effective action which 
contain only the effective gauge fields and involve an additional 
function $\cal{G}.$ We briefly described how regularization and 
renormalization are reflected in the dressing functions. The obtained 
effective action is assumed to contain the quantum contributions of the 
gauge theory, perturbative and non-perturbative, but not including 
the soliton-like vacuum effects.  Stated otherwise, all these effects are
assumed to be contained in a limited number  of dressing functions 
($G_A,G_c,\cal{G}$). Application and consistency checks of this 
effective action for the case of high-momentum QCD are presented elsewhere
\cite{second}.

\vskip 3mm 
\noindent {\large{\bf{Acknowledgements}}} 
\vskip 3mm

The work of I.K. was supported by the Programa Mecesup FSM9901 of the Ministry of
Education (Chile) and also by Conicyt (Chile) under grant 8000017. The work 
of G.C. and I.S. was supported by Fondecyt (Chile) grant No. 1010094. and  8000017, 
respectively. I.K. is grateful to Tim Jones for suggesting the terminology 
``dressing'' for the dressing functions.

\end{document}